\documentclass{epl}
\usepackage{graphicx,psfrag,amssymb,amsmath}
\title{Collective modes in unconventional density waves}
\shorttitle{Collective modes in unconventional density waves}
\author{Bal\'azs D\'ora\inst{1} \and Attila Virosztek\inst{2,3}}
\institute{
\inst{1} The Abdus Salam ICTP, P.O. Box 586, Trieste, I-34014, Italy\\
\inst{2} Department of Physics, Technical University of Budapest, H-1521 
Budapest, Hungary \\
\inst{3} Research Institute for Solid State Physics and Optics, P. O. Box
49,
H-1525 Budapest, Hungary}
\shortauthor{Bal\'azs D\'ora \etal}


\date{}

\begin{document}
\maketitle

\begin{abstract}
We have investigated the collective modes of unconventional charge and spin 
density waves (UCDW, USDW) 
in quasi-one dimensional systems 
in random phase approximation.
The density correlator regains its normal state form due to the phase degree of freedom of the condensate.
The possible effect of impurities is also discussed.
From this, the current-current correlation function is evaluated through charge conservation.
The spin susceptibility of USDW remains anisotropic an spite of the lack of any
periodic modulation of the spin density.
In UCDW, the spin response gets weaker in all three directions as the temperature is lowered.
\end{abstract}

 Unconventional density waves were intensively studied over the past few years. The so called
d-density wave, which is UCDW with a gap of d$_{x^2-y^2}$ symmetry, has attracted much attention\cite{nayak}.
This model successfully described many aspects of the pseudogap phase of high temperature cuprate superconductors.
Also d-wave SDW can account for the anomalously small magnetic ordering in URu$_2$Si$_2$\cite{IO,roma}.
In quasi-one dimensional systems, our UCDW model describes successfully the behaviour of the low temperature phase
of $\alpha$-(BEDT-TTF)$_2$KHg(SCN)$_4$\cite{rapid,tesla,imperfect}, including the angular dependent magnetoresistance\cite{alfa}.
The response of (TMTSF)$_2$PF$_6$ for $T<T^*\sim 4$K can be described
if the creation of USDW is assumed on top of the existing SDW \cite{basleticprb}.

The effect of collective modes is very important to study the dynamics of a
system. 
The pole structure of the
susceptibility is of prime importance\cite{He31,He32,caroli}: at a 
given combination of the
wavevector and frequency, the response function can be divergent. This
means, that even without external field, excitations occur in the system 
whose dispersion is determined by the poles of the susceptibility. One way
of identifying the correct order parameter among possible candidates is to
study the unique collective modes supported by the ground state of a given
symmetry\cite{p-wavecoll1,p-wavecoll2,dspin1,dspin2,higas}, as it was done in
unconventional superconductors. 

Allowing for a low temperature phase transition, the Green's function of
 the model changes and new type of anomalous Green's functions can enter
 (for example the Gor'kov type $F(i\omega_n,{\bf k})$ function belonging
 to the $\langle a^+_{\bf k,\downarrow}a^+_{-\bf k,\uparrow}\rangle$ pair
 correlation in superconductivity). Consequently new types of interaction
 channels might open due to the novel type of nonvanishing expectation
 value, which is called the fluctuation of the order parameter\cite{He31,He32}. 
As a result, the simple RPA series is modified,
 and a number of coupled geometrical series are to be summed up. 
In the case of a conventional DW, the collective modes are
 well-known\cite{Poilblanc1,Poilblanc2,viro3,viro5,viro6,CDW}. 
Generally one can conjecture whether the effect of RPA is or
 is not to re-establish the original pole structure of the metallic
 state. This can be achieved due to the degree of freedom of the phase of
 the density wave or the direction of the spin polarization of SDW. If we 
face with a magnetic phase transition (with a
 given spin orientation, i.e. SDW),
 the transverse (to the preferred direction) spin susceptibility and the 
density correlation function
 are not relevant from the transition's point of view, hence they will be
 restored. The former does due to the low energy (gapless) excitations of
 the spins in the perpendicular directions while the latter due to the free
 phase. Similarly the density correlator would regain its
 original form after RPA in a CDW due to the unrestricted phase of the
 density wave, but through the phonon mediated interaction the collective
 contribution is modified because of the mass enhancement. The quantities
 which are expected to
 change are the longitudinal spin susceptibility in SDW and all the spin
 susceptibilities in CDW because there are no low energy excitations based
 on the degrees of freedom of the system.

 From now on,
we introduce the formalism necessary to investigate the collective modes in
RPA and evaluate possible excitations of the main quantities.

\section{Formalism for RPA}

To start with, it is useful to introduce the spinor, which covers the whole momentum-spin space:
\begin{equation}
\Psi({\bf k}, \tau)=\left(\begin{array}{c}
      a_{{\bf k},\uparrow}(\tau) \\
      a_{\bf k-Q, \uparrow}\\
      a_{{\bf k},\downarrow}(\tau)\\
      a_{\bf k-Q, \downarrow}
\end{array}
\right),
\end{equation}
where $a_{\bf k,\sigma}$ is the annihilation operator of an electron of momentum $\bf k$ and
spin $\sigma$, $\bf Q$ is the best nesting vector.
From this the Green's function of USDW is obtained as
\begin{equation}
G({\bf k},i\omega_n)=-\frac{i\omega_n+\xi({\bf k})\rho_3+\Delta({\bf k})\rho_1\sigma_3}{\omega_n^2+\xi({\bf k})^2+\Delta({\bf k})^2},
\end{equation}
for UCDW $\sigma_3$ has to be replaced with $1$, where $\sigma_i$
($i=1,2,3$) are the Pauli matrices acting on spin space.
With this, the interaction responsible for the UDW formation is determined from the
gap equation of Ref. \cite{nagycikk}
and assuming $\Delta({\bf k})=\Delta\sin(bk_y)$ without loss of generality,
it is given by
\begin{gather}
\frac NV \tilde V({\bf
k,k^\prime,q},\sigma,\sigma^\prime)=\delta_{-\sigma,\sigma^\prime}(2J_y\sin(bk_y)\sin(b(k_y^\prime-q_y))-\nonumber \\ 
-2F_y\sin(bk_y)\sin(bk_y^\prime))+\delta_{\sigma,\sigma^\prime}(J_y-V_y)\sin(bk_y)\sin(b(k_y^\prime-q_y))
,\label{rpa1}
\end{gather}
Of course the
interaction is able to support $\Delta\cos(bk_y)$ type of gap, but we neglected
the terms favouring this specific wavevector dependence: they are irrelevant with respect to RPA because they can only renormalize the
coefficients of the susceptibilities but are unable to drive the system
into the desired ground state, namely with sinusoidal gap. This
approximation was used in the first step of the calculation, namely in the
gap equation\cite{nagycikk}, when the specific wavevector dependence of the gap was
chosen, because the matrix elements favouring this ordering are assumed to be
the strongest.  All the following calculations apply
also to a cosinusoidal gap. For simplicity we shall limit our analysis to 
${\bf q}=(q_x,0,0)$ (i.e. wavevector pointing in the quasi-one dimensional 
direction).

\section{Density correlator and complex conductivity}

For the density-density correlator we obtain for UDW 
\begin{eqnarray}
\langle[n,n]\rangle=\langle[n,n]\rangle_0+\frac{P_i}{4}\langle[n,A_i]\rangle_0\langle[A_i,n]\rangle,\\
\langle[A_i,n]\rangle=\langle[A_i,n]\rangle_0+\frac{P_i}{4}\langle[A_i,A_i]\rangle_0\langle[A_i,n]\rangle,
\end{eqnarray}
where $\langle\dots\rangle_0$ means the thermal average without interaction
between fluctuations. Here $i=c$, $s$ for UCDW and USDW,
respectively. $A_s=\rho_2\sigma_3\sin(bk_y)$, $A_c=\rho_2\sin(bk_y)$ and both $P_c$ and $P_s$
are positive. The detailed form of the interaction ($P$)  can be found in Ref. \cite{nagycikk}.
 The retarded products 
$\langle[n,n]\rangle_0$, etc. are
evaluated within the standard method and we find
\begin{subequations}
\begin{eqnarray}
\langle[n,n]\rangle_0&=&2g(0)\frac{\xi^2}{\xi^2-\omega^2}\left(1-
4\Delta^2F\right),\\
\langle[n,A_i]\rangle_0&=&-i4g(0) \xi\Delta F,\\
\langle[A_i,A_i]\rangle_0&=&2g(0)\left(\frac{2}{P_i g(0)}+(\omega^2-\xi^2)F\right)\label{pinning1},
\end{eqnarray}
\end{subequations}
where $g(0)$ is the density of states at the Fermi energy in the normal state per spin, $\xi=v_F q_x$, $\langle[n,n]\rangle_0({\bf q},\omega)$ is obtained, for
example, after
analytic continuation from
\begin{equation}
\langle[n,n]\rangle_0({\bf q},i\omega_\nu)=-\frac{1}{\beta V}\sum_{{\bf k},n}\textmd{Tr}(G({\bf
k},i\omega_n)G({\bf k-q},i\omega_{n-\nu}))
\label{suruseg}
\end{equation}
and
\begin{gather}
F=(\xi^2-\omega^2)\frac{1}{2\pi}\int\limits_0^\infty\int\limits_0^{2\pi}\tanh\frac{\beta
E}{2}\frac{N}{D}\textmd{Re}\frac{\sin(y)^2}{\sqrt{E^2-\Delta^2\sin(y)^2}}dydE,\\
N=(\xi^2-\omega^2)^2-4E^2(\xi^2+\omega^2)+4\Delta^2\sin(y)^2\xi^2,\\
D=N^2-64E^2\omega^2\xi^2(E^2-\Delta^2\sin(y)^2),
\end{gather}
is the $F$ function which also appears in the correlation functions of
conventional DW with constant gap\cite{viro5}. The extra $\sin(y)^2$ factor in the
numerator of the density of states like term comes either from the gap or
from the interaction. The remaining angle integral (which is
the Fermi surface average) can be performed and after straightforward manipulation it yields to
\begin{gather}
4\Delta^2
F=\frac{\xi^2-\omega^2}{\xi^2}\frac{2}{\pi}\left\{\int\limits_0^\Delta
dE\tanh\frac{\beta
E}{2}\left[\frac{N_1}{2\Delta D_1}\Pi\left(-\frac{4\xi^2E^2}{D_1},\frac
E\Delta\right)+\frac 1\Delta K\left(\frac E\Delta\right)+\right.\right.\nonumber \\
\left.\left.+\frac{N_2}{2\Delta
D_2}\Pi\left(-\frac{4\xi^2 E^2}{D_2},\frac
E\Delta\right)\right]\right.\nonumber+ \\
+\left.\int\limits_\Delta^\infty dE \tanh\frac{\beta
E}{2}\left[\frac{N_1}{2E D_1}\Pi\left(-\frac{4\xi^2\Delta^2}{D_1},\frac
\Delta E\right)+\frac 1 EK\left(\frac\Delta E\right)+\frac{N_2}{2E
D_2}\Pi\left(-\frac{4\xi^2\Delta^2}{D_2},\frac\Delta E\right)\right]\right\},
\end{gather}
where $K(z)$ and $\Pi(n,z)$ are the complete elliptic integral of the first and third kind\cite{elliptic}, respectively and
\begin{subequations}
\begin{eqnarray}
N_1&=&(-E^2-\xi^2+(E-\omega)^2)(\xi^2-(2E-\omega)^2),\\
D_1&=&(E^2+\xi^2-(E-\omega)^2)^2-4\xi^2E^2,\\
N_2&=&(-E^2-\xi^2+(E+\omega)^2)(\xi^2-(2E+\omega)^2),\\
D_2&=&(E^2+\xi^2-(E+\omega)^2)^2-4\xi^2E^2.
\end{eqnarray}
\end{subequations}
Putting these expressions together we obtain
\begin{equation}
\langle[n,n]\rangle(\omega,q)=2g(0)\frac{\xi^2}{\xi^2-\omega^2},
\end{equation}
 which is the same as in the normal state. It is worth mentioning that
in conventional SDW the coefficient of $\xi^2$ in the denominator is
$1+Ug(o)$\cite{viro5}, $U$ is the on-site Coulomb repulsion. The strength of interaction is missing here due to its zero
average over the Fermi surface (See Eq. (\ref{rpa1})).
In the absence of the
DW pinning, all the modifications due to the change of quasiparticle
spectrum are exactly canceled from the contribution of the fluctuation of
the order parameter. The coupled RPA equations consist of two parts: the
single particle contribution (one bubble contribution) which arises from the thermal excitation
across the gap and possibly from a remnant portion of the Fermi surface, and
the collective part from the motion of the density wave as a whole (due to
the freedom of the phase)\cite{rice}. In
the case of $\langle[n,n]\rangle$, these two cancel each other to give back
the normal state form. 

In real systems, impurities are always
present. Henceforth the easiest way to incorporate the effect of impurities
is to modify Eq. (\ref{pinning1}) as in Ref \cite{viro5} as
\begin{eqnarray}
\langle[A_i,A_i]\rangle_0=2g(0)\left(\frac{2}{P_i
g(0)}-(\xi^2+\omega_p^2-\omega^2)F\right),
\end{eqnarray}
where $\omega_p$ is the pinning frequency. This is the zeroth order phason
propagator, and the inclusion of $\omega_p$ ensures the presence of gapped
phason mode. In general, impurities act differently on UDW\cite{ohashi} 
than on conventional DW, but extended impurities are able to pin the phase 
of UDW, inducing finite pinning frequency\cite{rapid}.
As a result, the density correlator reads as
\begin{eqnarray}
\langle[n,n]\rangle(\omega,q)=2g(0)\xi^2\frac{\xi^2-\omega^2+\omega_p^2(1-f)}{(\xi^2-\omega^2)(\xi^2+\omega_p^2-\omega^2)}
\end{eqnarray}
for both USDW and UCDW, $f=4\Delta^2 F$. The zero sound dispersion ($\omega^2=\xi^2$)
cancels out here contrary
to the conventional case \cite{viro5}, because a
detailed study of the $f$ function reveals that
$f=1+\textmd{const}(\xi^2-\omega^2)/\Delta^2$ in the limit of $\omega$
tends to $\xi$.  
The pole of the propagator describes the pinned dynamics of the USDW and UCDW
condensate as $\omega^2=\omega_p^2+\xi^2$.
Due to pinning, the condensate does not move below the threshold frequency
$\omega_p$. 
The sound velocity in the presence of the electron-phonon coupling ($g_{e-ph}$) is given
by
\begin{equation}
C=C_0\sqrt{1-g_{e-ph}^2\lim_{q\rightarrow 0}\langle[n,n]\rangle(0,q)},
\end{equation}
where $C_0$ is the sound velocity without the electron-phonon coupling.
First, in the absence of the DW pinning the sound velocity is temperature
independent and reads as
\begin{equation}
C=C_0\sqrt{1-\lambda}
\end{equation}
with $\lambda=2g_{e-ph}^2g(0)$. 
Second, in the presence of pinning we obtain for UDW
\begin{equation}
C=C_0\sqrt{1-\lambda(1-\rho_s)},
\end{equation}
where $\rho_s$ is the static condensate density\cite{nagycikk}, $\rho_s=\lim_{q\rightarrow 0}\lim_{\omega\rightarrow 0}f$.
Since $\rho_s$ increases as the temperature decreases, we
predict increasing sound velocity by lowering the temperature below its
critical value $T_c$ for both UDW system. 

From the density correlator the complex conductivity in the chain
direction is obtained using charge conservation as
\begin{eqnarray}
\sigma(\omega,q)=2g(0)ie^2\omega\frac{\omega^2-\omega_p^2(1-f)-\xi^2}{(\xi^2-\omega^2)(\xi^2+\omega_p^2-\omega^2)}.
\end{eqnarray}
From this the optical conductivity of UDW reads as
Re$\sigma(\omega)=2g(0)e^2\pi\delta(\omega)$ in the pure case, which is essentially the same
formula as in the normal state. Note that this formula applies for UCDW
because there is no mass renormalization associated with lattice
distortion. If conventional CDW is deduced from the attractive Hubbard
model\cite{rice,viro3}, its conductivity coincides with those of a
conventional SDW (using a
phonon mediated attractive interaction, the mass gets renormalized in CDW).
In the $q\rightarrow 0$ limit the conductivity is the same as in
conventional SDW, only the $f$ function should be replaced by its
version corresponding to UDW:
\begin{equation}
\sigma(\omega,0)=2g(0)\frac{e^2}{i\omega}\left(f_0-1-\frac{f_0\omega^2}{\omega^2-\omega_p^2}\right)
\end{equation}
with $f_0=4\Delta^2F_{q=0}$. It approaches the one in the normal state for
$\omega\gg\omega_p$ for USDW and UCDW, while this statement is not true in
CDW because $m^*/m\gg1$. The peak at zero frequency in the optical conductivity moves to the
pinning frequency, suggesting that the primitive mimicking of pinning was
 successful.

The conductivity for electric field applied in the $z$ direction remains unchanged
because the interaction is unable to dress the single bubble contribution
due to the wavevector dependence of the velocity. In the $y$ direction
where the gap is developed, the following equation is to be solved:
\begin{equation}
\langle[j_y,j_y]\rangle=\langle[j_y,j_y]\rangle_0-\frac{P_j}{4e^2v_y^2}\langle[j_y,j_y]\rangle_0\langle[j_y,j_y]\rangle,  
\end{equation}
where $P_j=-P_c$, the matrix element responsible for the UCDW instability, and the current correlator without interaction
between fluctuations is obtained as
\begin{equation}
\langle[j_y,j_y]\rangle_0(\omega,q)=4e^2v_y^2g(0)\frac{1}{\xi^2-\omega^2}(\xi^2-\omega^2 f),
\end{equation}
which gives the paramagnetic part of the total conductivity. Here the
explicit wavevector dependence of the gap plays an important role. If the
system possess a gap with cosine, the interaction is unable to dress the
one bubble diagram, hence it gives the total paramagnetic part. The RPA
equation is true only for a gap $\sim\sin(bk_y)$.  
Adding the diamagnetic term to it, the complex conductivity reads as
\begin{equation}
\sigma_{yy}=\frac{\sigma_{yy0}-\dfrac{g(0)
P_j}{2}(\sigma_{yy0}-\sigma_n)}{1-\dfrac{g(0)
P_j}{2}\left(\dfrac{\sigma_{yy0}}{\sigma_n}-1\right)},
\end{equation}
where $\sigma_n=-2g(0)e^2v_y^2/i\omega$ is the normal state
conductivity, $\sigma_{yy0}$ is the conductivity without RPA. Only Fermi liquid type
corrections are present. 
Hence the excitation spectrum in the $y$ direction is given by
$\omega^2=(1+g(0) P_j (1-f_0))\xi^2$.
This is very similar to the zero sound dispersion in the $x$ direction in
conventional SDW in the presence of pinning\cite{viro5}. The sign of $P_j$ is negative
for UCDW but for USDW it can be positive as well. These calculations verify our earlier
assumption in Ref. \cite{nagycikk}, that the optical conductivity in the
perpendicular direction is given by the quasiparticle contribution, only
Fermi liquid renormalization may occur.  

\section{Spin susceptibilities}

As to the spin susceptibility, the transverse one in USDW is expected to
return to its normal state form similarly to the conventional case and to
exhibit the trivial Goldstone mode. The RPA
equations read as
\begin{eqnarray}
\langle[\sigma_1,\sigma_1]\rangle=\langle[\sigma_1,\sigma_1]\rangle_0+\frac{P}{4}\langle[\sigma_1,\rho_1\sigma_1]\rangle_0\langle[\rho_1\sigma_1,\sigma_1]\rangle,\\
\langle[\rho_1\sigma_1,\sigma_1]\rangle=\langle[\rho_1\sigma_1,\sigma_1]\rangle_0+\frac{P}{4}\langle[\rho_1\sigma_1,\rho_1\sigma_1]\rangle_0\langle[\rho_1\sigma_1,\sigma_1]\rangle,
\end{eqnarray}
where 
\begin{eqnarray}
\langle[\sigma_1,\sigma_1]\rangle_0=\frac{2g(0)}{\xi^2-\omega^2}(\xi^2-4\Delta^2\omega^2F),\\
\langle[\sigma_1,\rho_1\sigma_1]\rangle_0=4g(0)i\Delta\omega F,\\
\langle[\rho_1\sigma_1,\rho_1\sigma_1]\rangle_0=2g(0)\left(\frac{2}{g(0)
P}+(\omega^2-\xi^2)F\right).
\end{eqnarray}
Putting these together we obtain
\begin{equation}
\langle[\sigma_1,\sigma_1]\rangle=2g(0)\frac{\xi^2}{\xi^2-\omega^2}.
\end{equation}
In the static limit, this expression reduces to the well known Pauli susceptibility.
The longitudinal susceptibility of USDW shows only Fermi liquid type renormalization,
hence it is given mainly by the one bubble contribution, which coincids with the expression 
found for $\langle[n,n]\rangle_0({\bf q},i\omega_\nu)$ in Eq. (\ref{suruseg}). 
For UCDW the spin response is described with this expression in either
direction. Although there is no obvious magnetic long range order in USDW\cite{nagycikk},
yet it retains the spin anisotropy of SDW: in the static, long wavelength
limit the longitudinal susceptibility reads as
$\langle[\sigma_3,\sigma_3]\rangle=1-\rho_s$, which vanishes as $T$ goes to zero.
The pole belonging to the usual spin wave dispersion $\omega^2=\xi^2$
cancels out similarly to the case of the density correlator in the presence of pinning.

\section{Conclusion}
We have studied the relevant correlation functions with random phase
approximation. Again as in superfluid $^3$He\cite{He31,He32}, the 
quantity whose correlator is investigated, often couples to the fluctuation
of the UDW order parameter and it is necessary to handle these fluctuations
on equal footing. Due to the unrestricted phase of the density wave, the
density correlator regains its simple normal state form as it does in
conventional spin density waves. 
 In UCDW, the phason mass remains unrenormalized because
the interaction from which this phase is originated, is electron-electron
interaction and there are no retardation effects associated with
electron-phonon interaction\cite{CDW,viro3}. Based on a very simple concept\cite{viro5} the pinning of UDW
is incorporated into the theory and the sound velocity $C$ is examined in the
presence and absence of the pinning. Without pinning $C$ does not change
compared to the normal state value. In a pinned UDW the sound velocity
increases upon entrance in UDW. Additionally the conductivity in the chain
direction is obtained from the density correlator through charge
conservation. It does neither change compared to the one in the normal
state: in the pure system Re$\sigma(\omega)\sim\delta(\omega)$, which
inspired the first pioneers in this field to identify the origin of
the mechanism leading to superconductivity as CDW formation.

By evaluating the current-current correlation function in the perpendicular
directions, our previous assumption in Ref. \cite{nagycikk} is verified:
collective contributions do not show up and the quasiparticle contribution
is enough to describe the electromagnetic response of the UDW.

The spin susceptibility of USDW shows antiferromagnetic like
ani\-so\-tro\-py\cite{Chen,johnston} in spite of the lack of magnetic ordering, while in UCDW the
magnetic response is influenced mainly by the quasiparticle contribution,
leading to the freezing out of the static susceptibilities at low
temperatures in all directions, as we predicted in Ref. \cite{nagycikk}.
\begin{acknowledgments}
This work
was supported by the Hungarian National Research Fund under grant numbers
OTKA T032162 and  TS040878, and by the Ministry of Education under grant 
number FKFP 0029/1999.
\end{acknowledgments}

\bibliographystyle{apsrev}
\bibliography{eth}
\end{document}